# Automated Keypoint Estimation for Self-Piercing Rivet Joints Using μCT Imaging and Transfer Learning

Wei Qin Chuah
*School of Engineering*
*RMIT University*
Melbourne, Australia
wei.qin.chuah@rmit.edu.au

Ruwan Tennakoon
*School of Computing Technologies*
*RMIT University*
Melbourne, Australia

Amanda Freis
*Ford Research and Innovation Center,*
*Ford Motor Company,*
Dearborn, MI 48124, USA

Mark Easton
*School of Engineering*
*RMIT University*
Melbourne, Australia
mark.easton@rmit.edu.au

Reza Hoseinnezhad
*School of Engineering*
*RMIT University*
Melbourne, Australia
rezah@rmit.edu.au

Alireza Bab-Hadiashar
*School of Engineering*
*RMIT University*
Melbourne, Australia
abh@rmit.edu.au

*Abstract*—The structural integrity of self-piercing rivet (SPR) joints is critical in automotive industries, yet its evaluation poses challenges due to the limitations of traditional destructive methods. This research introduces an innovative approach for non-destructive evaluation using μCT imaging, Micro-Computed Tomography, combined with machine vision and deep learning techniques, specifically focusing on automated keypoint estimation to assess joint quality.

Recognizing the scarcity of real μCT data, this study utilizes synthetic data for initial model training, followed by transfer learning to adapt the model for real-world conditions. A UNet-based architecture is employed to localize three keypoints with precision, enabling the measurement of critical parameters such as head height, interlock, and bottom layer thickness. Extensive validation demonstrates that pre-training on synthetic data, complemented by fine-tuning with limited real data, bridges domain gaps and enhances predictive accuracy.

The proposed framework not only offers a scalable and cost-efficient solution for evaluating SPR joints but also establishes a foundation for broader applications of machine vision and non-destructive testing in manufacturing processes. By addressing data scarcity and leveraging advanced machine learning techniques, this work represents a significant step toward automated quality control in engineering contexts.

## I. Introduction

Self-piercing rivets (SPRs) are extensively used in the vehicle manufacturing industry because of their superior mechanical properties and their ability to join dissimilar materials without requiring pre-drilled holes [1]. During riveting, SPR applies compressive stress around the joint, which helps prevent crack initiation and propagation, further enhancing joint integrity [2]. Ensuring the quality of SPR joints is critical for maintaining structural integrity in these applications [3]. Accurate assessment techniques are vital to detect potential defects or inconsistencies in the joints [4]. Traditional evaluation methods, such as destructive testing, are both costly and impractical for large-scale operations, leading to an increased interest in non-destructive evaluation (NDE) methods. Among these, μCT imaging has emerged as a promising alternative, offering detailed insights into the internal structure of SPR joints [5, 6].

In this study, an innovative framework that automates the evaluation of SPR joints using μCT imaging combined with deep learning techniques [7] is proposed. By focusing on keypoint estimation, the framework provides a precise tool to measure critical structural features, automating quality assessments. However, the availability of μCT data for SPR joints is limited due to the high cost and complexity of acquiring such data. This data scarcity hinders the development of reliable models for automated keypoint estimation.

To overcome this challenge, the proposed approach leverages synthetic data, which are abundant and easily produced using simulation software, for initial model training. However, due to significant distributional differences from real μCT images, models trained solely on synthetic data perform poorly on real μCT images. To bridge this gap, transfer learning is employed, where a model pre-trained on synthetic data is fine-tuned with minimal real μCT data, resulting in a robust and generalizable keypoint [8-10] estimation framework. This approach enables the development of a robust keypoint estimation model capable of achieving high accuracy with limited real data.

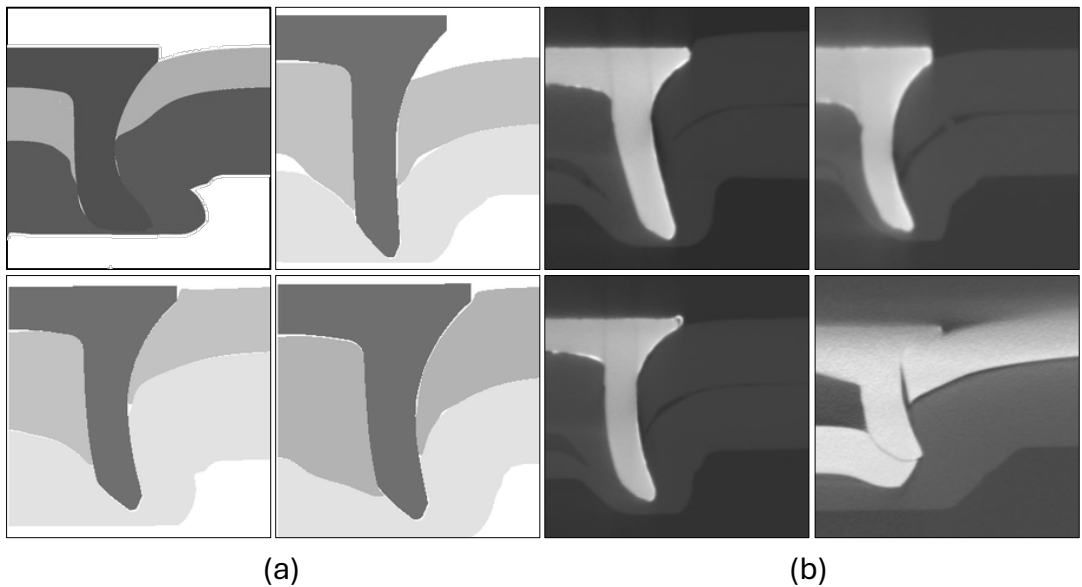

Fig. 1. Qualitative comparison of SPR joints between (a) synthetic data and (b) real μCT data.

The proposed framework includes detailed steps for generating, preprocessing and annotating synthetic data and real μCT data, as well as fine-tuning the model specially designed for the task. By integrating these methodologies, this study aims to establish a non-destructive, cost-effective and scalable solution for estimating keypoints in SPR joints with high precision, which in turn could be used for evaluating the joint quality.

## II. DATASET DESCRIPTION

This study leverages both synthetic and real μCT datasets to develop a robust keypoint estimation model for Self-piercing rivet (SPR) joints. Synthetic data, generated using Simufact® simulation software, offers a clean and controlled training environment but lacks real-world nuances, such as material heterogeneity and unavoidable variations in manufacturing conditions and component properties. Real μCT data, derived from SPR joints fabricated with varying materials, adhesives, and rivet head configurations, address these complexities, enhancing the model's generalization.

Figure 1 illustrates a qualitative comparison of synthetic (a) and real μCT data (b), highlighting the differences in complexity and realism between the two datasets. These steps collectively form the foundation for accurate keypoint localization and quality assessment.

The following sections detail the synthetic and real data characteristics, along with the preprocessing and labelling steps undertaken to prepare the dataset for this task.

### A. Synthetic Data

Synthetic data of the SPR joints were generated using the Simufact simulation software, offering a controlled and noise-free environment for initial model training. Widely employed for simulating the SPR process via non-linear finite element analysis across diverse aluminium alloy sheets, the Simufact simulation software has effectively demonstrated its capability by accurately modelling the specific material combinations and gauges examined in this study [12]. This dataset comprises six rivet types paired with two material types: DC04 steel sheets and A6061-T6 aluminium alloy sheets. Thickness values were systematically selected from the set: 1.0, 1.2, 1.5, 1.6, 1.8, 2.0, 2.2, and 2.5 mm. For example, one configuration might involve a top sheet of 1.2 mm and a bottom sheet of 2.0 mm, with both sheets composed of the same material to ensure consistency across simulations.

While synthetic data provide a clean and structured training environment, they lack the nuances of real-world data, such as material heterogeneity, manufacturing variation, and scanning artifacts. These limitations can hinder direct generalization to practical applications. For example, optimizing machine learning models solely using the synthetic data may result in models that are over-specialized to these data. Nevertheless, synthetic data excel in facilitating the initial training phase, enabling the model to identify critical patterns of the keypoints under controlled conditions. Keypoints in synthetic images were manually annotated and subsequently converted into spatial heatmaps to serve as dense, informative labels for supervised learning. The detailed labelling process is elaborated upon in the preprocessing section (see below).

### B. Real Data

To overcome the limitations of synthetic data, real μCT data were incorporated to reflect the variability and imperfections characteristic of real-world conditions. These data were derived from μCT scans of self-piercing rivet (SPR) joints fabricated using three materials sheet types: AA5754-O, cold-rolled DP800, AlSi10MgMn-T7. To introduce additional variability, two adhesive types (DuPont Betamate 6160 and Sika Power 510G) and three rivet head height configurations (nominal, underflush, and overflush) were included, resulting in 45 distinct SPR configurations.

Each joint underwent comprehensive 3D μCT scanning, and 2D slices were extracted at 90-degree rotational intervals, yielding 180 images in total. Unlike their synthetic counterparts, real μCT images feature intrinsic imperfections such as manufacturing variations, material heterogeneity, and scanning artifacts. These challenges, while complicating keypoint estimation, are critical for ensuring the model's robustness and applicability in real-world contexts. By fine-tuning models trained on synthetic data with real μCT data, the inherent distributional disparities were effectively bridged, enhancing generalization and predictive accuracy.

## III. Preprocessing

Preprocessing played a pivotal role in standardizing the dataset and optimizing it for model training and evaluation. To ensure unbiased assessment, the dataset was split into training and testing subsets with an 8:2 ratio. Care was taken to avoid overlap between subsets, ensuring that data from a specific SPR configuration were exclusive to either the training or testing set. This strategy mitigated the risk of data leakage, preserving the integrity of the evaluation process.

For both synthetic and real μCT datasets, images were cropped to isolate the region of interest, focusing exclusively on the rivet joint and eliminating irrelevant background information. The cropped images were then resized to a standardized resolution of 224×224, following established ImageNet preprocessing protocols. This resizing process carefully preserved the aspect ratio and structural integrity of the images to avoid distortions that could undermine keypoint localization accuracy. These preprocessing steps were essential for harmonizing the synthetic and real datasets, enabling seamless integration during training and evaluation.

### A. Labelling

The labelling process was a cornerstone of this study, ensuring the creation of accurate and reliable ground-truth annotations for supervised learning. The description of the ground truth keypoints will be discussed in the next section. Trained annotators identified key structural elements of the rivet joints and manually marked precise locations on each image. These annotations were then transformed into spatial heat maps, representing the probability of keypoint presence across the image. These heatmaps provided dense, high-quality target outputs for the convolutional UNet model, facilitating the learning of precise keypoint locations. Representative samples of synthetic and real annotated images, shown in Figure 2, illustrate the differences in data complexity and highlight the thoroughness of the labelling effort. This meticulous process ensured the creation of a dataset capable of supporting reliable and accurate keypoint estimation across a wide spectrum of scenarios.

### B. Keypoints Description and Critical SPR Measurements

Figure 2 illustrates a qualitative comparison between synthetic SPR joint data (left) and real μCT data (right), highlighting three critical measurements: head height ($d_h$), interlock ($d_i$), and minimum bottom thickness ($d_b$), using six keypoints. These measurements are visualized to provide insight into the quality and structural integrity of SPR joints. The keypoints are color-coded and paired to correspond to the specific measurements: red and yellow for $d_h$, green for $d_i$, and blue for $d_b$.

The **head height ($d_h$)**, computed using the top two keypoints (red and yellow), represents the vertical distance from the top edge of the SPR joint to the surrounding material surface. This parameter is crucial for assessing the penetration depth and the joint's repeatability.

The **interlock ($d_i$)**, derived from the middle keypoints (green), measures the lateral engagement of the material layers at the joint. This dimension reflects the mechanical interconnection between layers and is a key indicator of the joint's resistance to shear forces and out-of-plane loading, directly influencing its strength.

The **minimum bottom thickness ($d_b$)**, determined by the bottom keypoints (blue), defines the remaining material thickness at the base of the joint. This measurement is vital for ensuring that the joint retains sufficient material to prevent breakthrough in the joint to prevent corrosion in wet environments.

## IV. Network Architecture Design

The proposed framework utilizes a UNet-based convolutional architecture optimized for keypoint estimation in μCT images of SPR joints. The network follows an encoder-decoder design with skip connections to preserve spatial details during upsampling (the process of increasing the resolution or size of input data, typically images or feature maps, to match a desired output size). The encoder extracts hierarchical features through successive convolutions and downsampling, while the decoder reconstructs fine-grained spatial details, enabling precise localization of keypoints.

To enhance performance, the model outputs heatmaps representing the likelihood of keypoints across the image. Instance normalization and ReLU activation, Rectified Linear Unit: an activation function commonly used in neural networks, are incorporated for stability and non-linear feature extraction. The architecture effectively balances global context understanding with pixel-level accuracy, making it well-suited for machine vision applications in non-destructive evaluation.

### A. UNet Heatmap Regression Model

The network architecture employed in this study is a convolutional UNet-based model, specifically tailored for heatmap regression tasks to predict the keypoints of SPR joints in μCT images. The architecture follows a symmetric encoder-decoder design with key modifications for the unique requirements of SPR joint analysis. The schematics of the network design are shown in Figure 3, illustrating the flow of information and structural components.

### B. Encoder-Decoder Architecture with Skip Connections

The UNet model consists of two primary components: the encoder and the decoder, both of which are implemented using convolutional blocks with specific kernel sizes and strides. The encoder utilizes 3x3 convolutional layers with stride 2 (denoted as Conv2D, k=3, s=2) to downsample the input image progressively. The encoder is responsible for extracting hierarchical feature representations of the input image through successive convolutional layers and downsampling operations. This process captures the global context of the image while reducing its spatial dimensions. The decoder employs transposed convolutional layers (denoted as DeConv2D, k=3, s=2) to upsample the features back to the original resolution, ensuring precise spatial reconstruction. The decoder also integrates fine-grained details through skip connections from the encoder. These connections ensure that spatial information lost during downsampling is preserved and integrated into the upsampling process, enabling precise localization of keypoints. This design is particularly advantageous for keypoint estimation, where pixel-level accuracy is critical. The architecture also includes instance normalization to stabilize training and ReLU activations for non-linear feature extraction after every convolution operation, except the final layer.

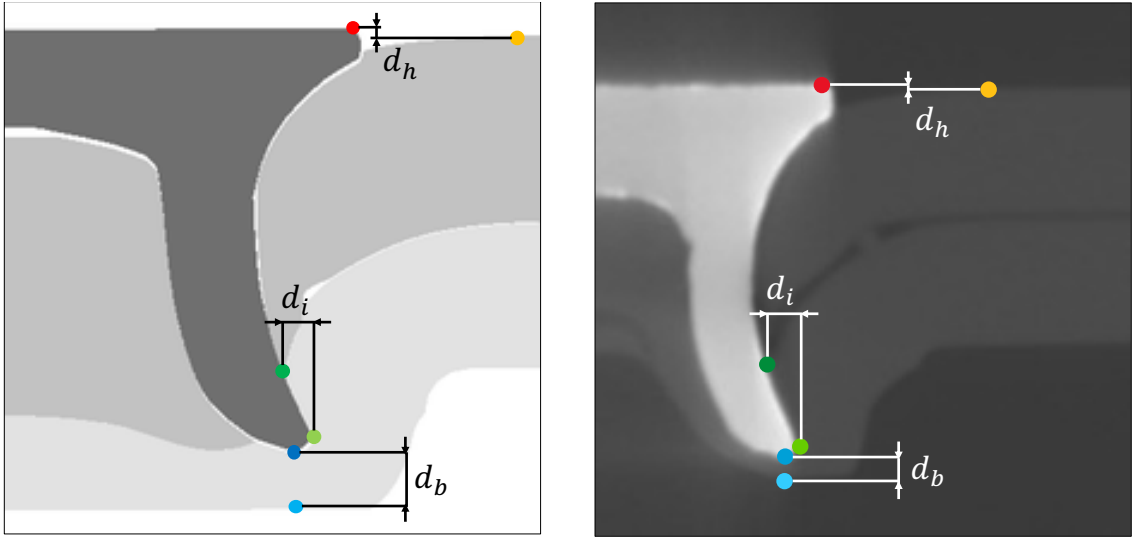

Fig. 2. Qualitative comparison of SPR joints: synthetic data (left) vs. real µCT data (right). Keypoints and corresponding measurements are illustrated to visualize critical joint parameters: head height ($d_h$), interlock ($d_i$), and minimum bottom thickness ($d_b$). Red and yellow keypoints are used for computing $d_h$, green keypoints for $d_i$, and blue keypoints for $d_b$. These parameters are essential for assessing the structural integrity and quality of SPR joints.

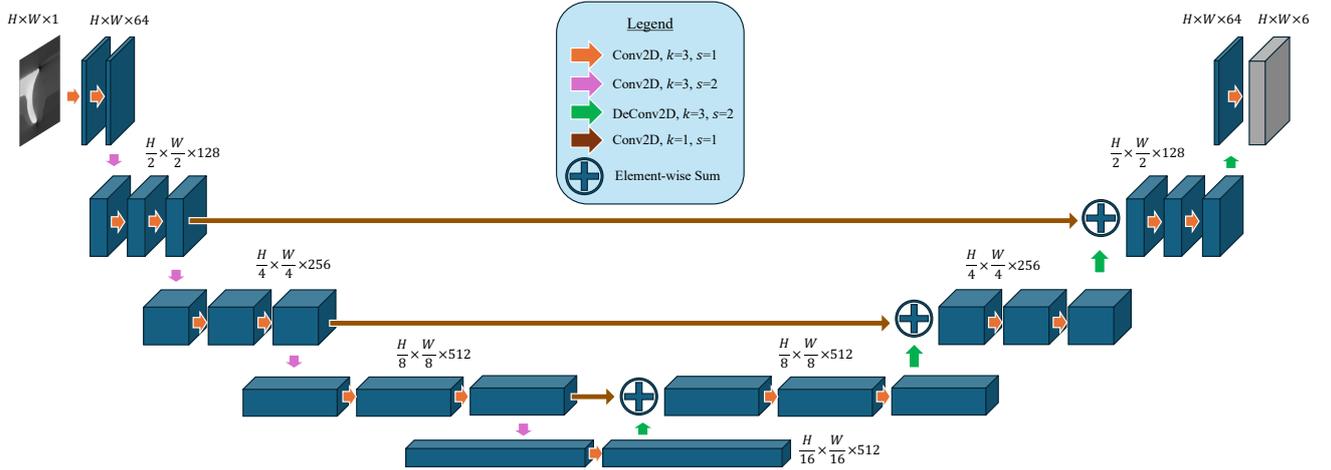

Fig. 3. Schematic of the UNet Heatmap Regression Model. The diagram illustrates the encoder-decoder architecture, highlighting 2D convolutional layers with varying kernel sizes (k=3, k=1) and strides (s=1, s=2), 2D transpose convolutional layers with kernel size 3x3 and stride 2, and element-wise summation layers. The model takes a preprocessed µCT image of an SPR joint as input, processes it through the encoder-decoder pathway, and outputs a set of heatmaps, each representing the likelihood of the presence of specific keypoints within the image.

To adapt the UNet model for keypoints estimation, the final layer of the decoder outputs a set of heatmaps, each corresponding to a specific keypoint. Each heatmap encodes the likelihood of the keypoint's presence at each pixel location. The model employs a softmax activation function to ensure that the output heatmaps represent valid probability distributions.

### C. Loss Function and Optimization

The training and fine-tuning process of the UNet model are guided by a pixel-wise binary cross-entropy loss function, which is defined as follows:

$$\mathcal{L}_{CE} = -\frac{1}{NM}\sum_{i=1}^{N}\sum_{j=1}^{M}\left[H_{ij}\log(\widehat{H}_{ij}) + (1 - H_{ij})\log(1 - \widehat{H}_{ij})\right] \quad (1)$$

where $H_{ij}$ and $\widehat{H}_{ij}$ represent the ground truth and predicted heatmap values at pixel $(i,j)$, respectively. The terms $N$ and $M$ denote the total number of pixels in the heatmap and the number of keypoints, respectively. The binary cross-entropy formulation treats each pixel as an independent binary classification problem, where $H_{ij}$ indicates the presence $(H_{ij} \approx 1)$ or absence $(H_{ij} \approx 0)$ of a keypoint at that location. This approach ensures a fine-grained focus on pixel-level precision, making it particularly effective for heatmap regression tasks for keypoint estimation where accurate keypoint localization is critical.

To generate the ground truth heatmaps, annotated keypoint locations are transformed into Gaussian distributions with an initial standard deviation (σ) of

3. This representation ensures a smooth and broad likelihood distribution, allowing the model to learn coarse patterns effectively during the initial stages of training.

The model undergoes a two-phase training process. First, pre-training is conducted on synthetic data for 100 epochs, allowing the model to develop a foundational understanding of keypoint localization. Fine-tuning is then performed for an additional 100 epochs using real μCT data. During the first 50 epochs of fine-tuning, the Gaussian standard deviation remains at σ = 3, emphasizing general spatial relationships. In the subsequent 50 epochs, σ is reduced to 1.5, sharpening the focus on precise keypoint localization. This two-stage approach enables the model to balance generalization and specificity, achieving robust and confident predictions even in the presence of noise and artifacts in real-world data.

To further enhance the model's robustness and generalization, data augmentation techniques are applied during both pre-training and fine-tuning stages. The applied data augmentation techniques include random affine transformations such as scaling (0.8–1.1), translations (±5 pixels), shear (±3°), and rotation (±2°), were utilized. A Gaussian blur with a randomly selected kernel size (5 or 7) and a fixed sigma of 20 was subsequently applied, and image intensities were clipped to the range [–1, 1]. By introducing variability into the training data, these augmentations help the model develop invariance to minor perturbations, ensuring consistent performance across diverse and noisy datasets.

The optimization process utilizes the Adam optimizer, renowned for its ability to adapt learning rates dynamically, which accelerates convergence while maintaining stability. The learning rate is initially set at 0.001, accompanied by a weight decay of $1\times10^{-6}$ to regularize the model and mitigate overfitting.

### D. Evaluation Metrics

To evaluate the performance of the developed model for keypoint estimation, three metrics are employed: Percentage of Correct Keypoint (PCK), Object Keypoint Similarity (OKS), and Mean Per-Joint Position Error (MPJPE). Each metric provides a unique perspective on the model's accuracy and robustness, ensuring a comprehensive assessment of its performance.

### E. Percentage of Correct Keypoint (PCK)

The PCK metric measures the percentage of keypoints predicted correctly within a specified distance threshold from their ground truth locations. A keypoint is considered correct if the Euclidean distance between the predicted and ground truth positions is less than a predefined threshold, typically a fraction of the object's size or image dimensions. Mathematically, PCK is expressed as:

$$\text{PCK} = \frac{1}{M}\sum_{i=1}^{M} \bar{1}(d_i \leq \tau) \qquad (2)$$

where $M$ is the total number of keypoints, $d_i$ is the Euclidean distance between the predicted and ground truth positions of the $i$-th keypoint, and $\bar{1}$ is an indicator function that equals 1 if the condition is met and 0 otherwise. Higher PCK values ($\text{PCK} \approx 1$) indicate better performance, demonstrating the model's capability to predict keypoints within acceptable tolerances. To specify the threshold used, it is common to denote the metric as PCK@τ. For example, PCK@10 indicates that the threshold is set to τ = 10 pixels.

### F. Object Keypoint Similarity (OKS)

The OKS metric evaluates the similarity between predicted and ground truth keypoints, considering both the spatial proximity and the scale of the object. It applies a Gaussian penalty to the distance between predicted and ground truth keypoints, normalized by the object scale and a constant factor. The OKS (Object Keypoint Similarity) is defined as [11]:

$$\text{OKS} = \frac{1}{M}\sum_{i=1}^{M} \exp\left(-\frac{d_i^2}{2s^2 k_i^2}\right) \qquad (3)$$

where $d_i$ is the Euclidean distance between the predicted and ground truth positions of the $i$-th keypoint, $s$ is the object scale, and $k$ is a constant controlling the Gaussian penalty for the $i$-th keypoint. More specifically, the constant parameter $k$ determines how quickly the penalty decreases as the distance between a predicted keypoint and its ground truth increases. A smaller $k$ value results in a sharper penalty curve, meaning even small deviations are penalized significantly. Conversely, a larger $k$ value creates a more gradual penalty, tolerating greater deviations. In the evaluation protocol of this work, $s$ is set as the rivet head radius and $k$ as 0.1. Similar to the PCK metric, a higher OKS score ($\text{OKS} \approx 1$) signifies that the predicted keypoints align well with their ground truth counterparts across different object scales.

### G. Mean Per-Joint Position Error (MPJPE)

The MPJPE metric quantifies the average Euclidean distance between predicted and ground truth keypoints across all joints. Unlike PCK, which uses a threshold-based criterion, MPJPE provides a direct measure of the model's overall precision. It is calculated as:

$$\text{MPJPE} = \frac{1}{M}\sum_{i=1}^{M} d_i \qquad (4)$$

where $d_i$ is the Euclidean distance between the predicted and ground truth positions of the $i$-th keypoint. Lower MPJPE values indicate better performance, as they reflect smaller deviations between predictions and ground truth locations. This metric is particularly useful for identifying areas where the model may struggle, offering insights into specific keypoints or joints that require further refinement.

## V. RESULTS AND DISCUSSION

### A. Quantitative Results

The quantitative results presented in the Table I, II and III demonstrate the critical role of synthetic data pre-training and fine-tuning in enhancing model performance for keypoint detection in SPR joints. The results highlight the varying performance of models across synthetic and real μCT datasets under different training configurations, providing valuable insights into the impact of pre-training, fine-tuning, and dataset limitations. In each of the tables, an up (↑) or down (↓) arrow is included next to the metric names to clearly indicate whether higher or lower values represent better performance.

*1) Performance on Synthetic Test Data*

The results for the pre-trained model on the synthetic test set included in Table I reveal high performance across all metrics, with PCK@10 and PCK@50 achieving 0.91 and 0.96, respectively, and an OKS of 0.91. These results indicate that the synthetic data captures the key geometric and structural features of SPR joints, allowing the model to generalize well to synthetic test samples. However, this success does not translate directly to real μCT data, as shown in subsequent experiments, due to the inherent differences in data characteristics between synthetic and real datasets.

*2) Performance on Real μCT Test Data*

TABLE I. QUANTITATIVE RESULTS OF A MODEL PRE-TRAINED ON SYNTHETIC DATA AND THEN EVALUATED ON OTHER SYNTHETIC TEST SET.

| PCK@10 (↑) | PCK@50 (↑) | MPJPE (↓) | OKS (↑) |
|---|---|---|---|
| 0.91 | 0.96 | 2.92 | 0.91 |

TABLE II. QUANTITATIVE RESULTS OF THE MODEL PRE-TRAINED ON SYNTHETIC DATA EVALUATED ON REAL **μCT TEST SET**.

| Model Variant | PCK@10 (↑) | PCK@50 (↑) | MPJPE (↓) | OKS (↑) |
|---|---|---|---|---|
| Pre-trained | 0.48 | 0.73 | 9.22 | 0.60 |
| Fine-tuned | 0.72 | 0.93 | 3.95 | 0.82 |

Table II compares the performance of two models on the real μCT test data: one pre-trained solely on synthetic data and the other fine-tuned on real μCT data. The pre-trained model achieves moderate performance on real data, with a PCK@10 of 0.48, PCK@50 of 0.73, and OKS of 0.60. This demonstrates the model's ability to transfer some knowledge from synthetic data to real-world conditions, but the domain gap limits its effectiveness. Fine-tuning the model on the real μCT training set significantly improves its performance, with PCK@10 increasing to 0.72, PCK@50 to 0.93, and OKS to 0.82. This improvement highlights the importance of bridging the domain gap through fine-tuning, as it enables the model to better align with the visual and structural complexities of real μCT data.

*3) Impact of Skipping Pre-training*

The results in Table III underscore the limitations of training directly on the real μCT dataset without synthetic pre-training. Due to the limited size of the real dataset, the model overfits, leading to poor generalization and significantly lower performance across all metrics. The PCK@10 and PCK@50 metrics drop to 0.13 and 0.37, respectively, with an OKS of 0.26 and an MPJPE of 34.89. These results clearly demonstrate that the small size of the real μCT dataset is insufficient to train a robust model, reinforcing the importance of synthetic data as a supplementary source for pre-training.

While increasing the amount of real μCT data could potentially address the issue of overfitting, acquiring additional real μCT data is often impractical due to the high cost and time-intensive nature of the process. In contrast, synthetic data offers a cost-effective and scalable alternative, enabling the generation of diverse and high-quality training samples without the significant expense or logistical challenges associated with real data collection. This makes the use of synthetic data a highly preferred approach for improving model robustness and generalization.

### B. Qualitative Results

The qualitative results presented in the Figure 4 demonstrate the impressive performance of the fine-tuned model on real μCT test data. The alignment between the predicted keypoints (dots) and the true keypoints (X marks) illustrates the model's ability to accurately localize critical features of the SPR joint, including head height (red and yellow keypoints), interlock (green keypoints), and bottom thickness (blue keypoints). The consistent proximity of predictions to ground truth locations across the different images highlights the robustness of the fine-tuned model, even in the presence of potential noise or variations inherent to real μCT data.

TABLE III. QUANTITATIVE RESULTS ON THE REAL μCT TEST SET USING A MODEL TRAINED DIRECTLY ON THE REAL μCT TRAINING SET, **WITHOUT PRE-TRAINING ON SYNTHETIC DATA**.

| PCK@10 (↑) | PCK@50 (↑) | MPJPE (↓) | OKS (↑) |
|---|---|---|---|
| 0.13 | 0.37 | 34.89 | 0.26 |

These results validate the effectiveness of the fine-tuning process, which bridges the domain gap between synthetic and real data, enabling the model to adapt to the visual and structural complexities of real-world SPR joints. The minimal deviation observed in most keypoints suggests that the fine-tuned model achieves a high degree of spatial accuracy, corroborating the quantitative metrics reported earlier, and provides visual evidence of the model's capacity to generalize and perform reliably on unseen real μCT data.

Apart from the predicted coordinates, examining the heatmaps could provide crucial insights into the model's confidence and focus during keypoint prediction. These insights offer a deeper understanding of its spatial reasoning and the reliability of its localization. By visualizing the distribution of confidence, the heatmaps allow for the assessment of not only the model's predictive accuracy but also its certainty, highlighting areas where it performs well and regions where it faces challenges. For instance, as shown in Figure 5, the heatmaps of keypoint 1, 4 and 6 display sharp, concentrated activations around the true keypoint locations, indicating high confidence and precise predictions. In contrast, keypoint 2, 3 and 5 exhibit broader and less-defined activations, suggesting greater uncertainty. This disparity likely arises from the increased geometric complexity or subtle visual features in the interlock region. A similar trend is observed in other μCT test data, with high confidence for keypoints 1, 4, and 6, while greater uncertainty is noted for keypoints 2, 3, and 5.

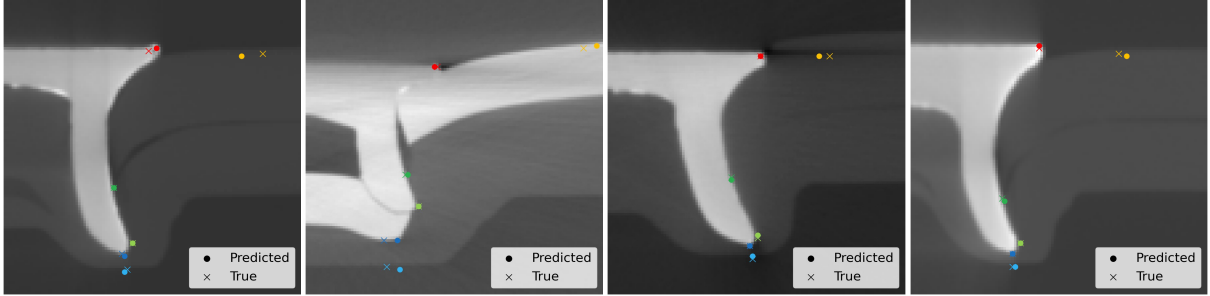

Fig. 4. Qualitative comparison of predicted and true keypoint locations on real μCT test data for the fine-tuned model. The colored "X" marks represent the true keypoint locations (red/yellow for head height, green for interlock, and blue for bottom thickness), while colored dot markers represent the predicted keypoints. The figure demonstrates the fine-tuned model's ability to closely align predictions with ground truth keypoints, highlighting its accuracy and robustness.

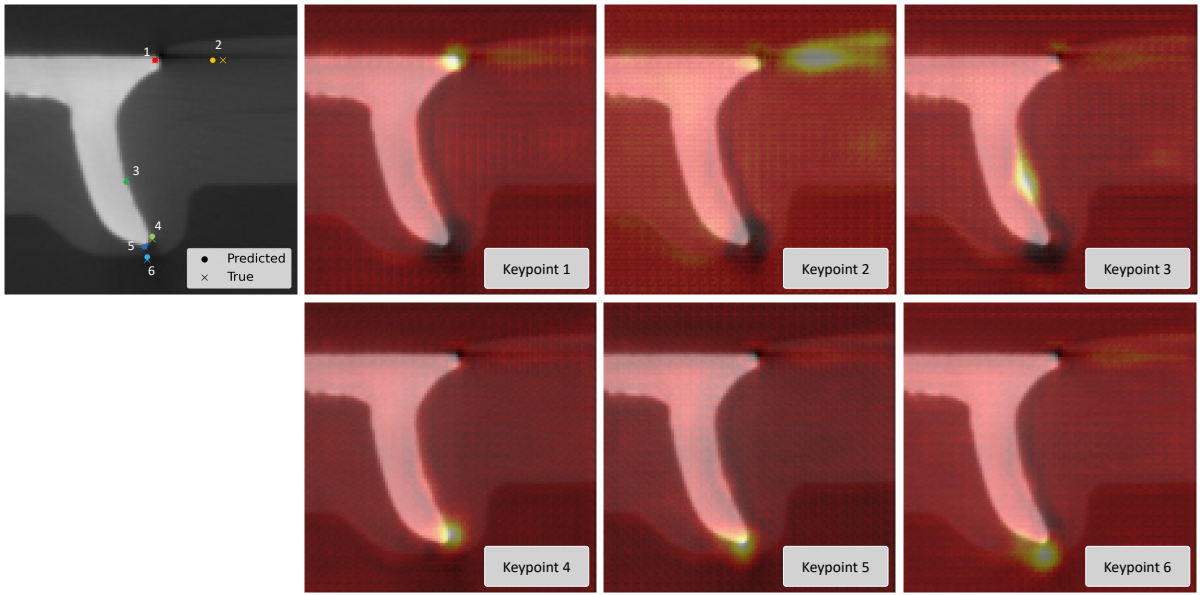

Fig. 5. Visualization of heatmaps of the model's confidence for each keypoint, with brighter regions indicating higher prediction certainty. The figure highlights the model's ability to accurately localize critical keypoints while providing insights into the spatial confidence distributions for each prediction.

## VI. CONCLUSION

This study introduces an innovative framework to estimate three keypoints in self-piercing rivet (SPR) joints using μCT imaging and deep learning methodologies. By integrating synthetic data pre-training with real μCT data fine-tuning, the framework effectively bridges the domain gap, achieving high accuracy and robustness in identifying the keypoints localization. The results demonstrate the significant advantage of leveraging synthetic data to overcome the limitations of scarce real-world data, emphasizing the practicality and scalability of the approach for non-destructive evaluation (NDE).

Quantitative metrics, including PCK, OKS, and MPJPE, highlight the model's strong performance, while qualitative visualizations corroborate its ability to handle noise and variability in real data. The incorporation of transfer learning and targeted fine-tuning techniques further reinforces the framework's capability to generalize across diverse conditions, making it a valuable tool for assessing the structural integrity of SPR joints.

This framework paves the way for advancements in automated quality control processes in engineering applications, offering a cost-effective and scalable solution. Future work will focus on enhancing model robustness through advanced data augmentation techniques and exploring alternative architectures to improve performance further.

**Acknowledgment**

This research was supported by the Australian Research Council (ARC) Linkage Projects under Grant LP190100165 and by the Ford Motor Company, Research and Innovation Center in Dearborn, USA, as the industry partner. The authors gratefully acknowledge their contributions and funding, which made this work possible.